\documentstyle[aps]{revtex}
\draft
\begin{document}
%
\title{Dephasing Effects by Ferromagnetic Boundary
on Resistivity in Disordered Metallic Layer
} 
\author{Gen Tatara}
\address{Max Planck Institut fur Mikrostrukturphysik
Weinberg 2, 06120 Halle,
Germany\\
and\\
Graduate School of Science, Osaka University, Toyonaka, 
Osaka 560-0043, 
Japan}
\author{Hidetoshi Fukuyama} 
\address{Department of Physics, University of Tokyo, Hongo, Tokyo 
113-0033, Japan}
\date{\today}
\maketitle
%
%
\begin{abstract}
The resistivity of disordered metallic layer sandwiched by two 
ferromagnetic layers at low-temperature is investigated theoretically. 
It is shown that the magnetic field acting at the interface does not 
affect the classical Boltzmann resistivity but causes a 
dephasing among electrons in the presence of the spin-orbit interaction, 
suppressing the anti-localization 
due to the spin-orbit interaction. 
The dephasing turns out to be stronger in the case where the magnetization 
of the two layers is parallel,
contributing to a positive magnetoresistance close to a 
switching field at low temperature.
\end{abstract}

In metallic magnets the electronic 
transport properties can strongly be affected by the configuration of the 
magnetization. 
Especially the resistivity close to the coercive field will vary by  
a small magnetic field due to a rearrangement of the 
magnetization.
This effect, called a magnetoresistance (MR), has been observed for a 
long time in bulk magnets since early time\cite{Thomson1857}.
The MR in bulk magnets is anisotropic, in the sense that it depends on 
the mutual angle 
between the current and the magnetization. 
The change of resistivity is of order of a few \% of the total 
resistivity.
This anisotropic MR is explained by the 
spin-orbit interaction in $d$-band\cite{McGuire75}.
In 1988 MR of about 50\% has been found in multilayer structures of 
Fe and Cr\cite{Baibich88}. 
Such MR seen in multilayers, which is called a giant 
MR (GMR),  is believed to be mainly due to the 
spin-dependent scattering of the electron at the interface\cite{Baibich88}.

Quite recently the MR in a mesoscopic magnetic 
structures has been studied intensively, for instance, on sub-micron 
wires of ferromagnetic 
metals\cite{Giordano94,Hong96,Otani97,Ruediger98,Ono98} 
and on a multilayer structure of a hard- and soft-magnets\cite{Mibu98}.
So far transport properties in such magnetic structures has 
mostly been discussed in terms of classical theories. 
For instance, the Boltzmann resistivity due to a reflection by domain wall 
has been calculated\cite{Cabrera74}, whose result indicated a negligiblly 
small contribution in 3$d$ transition metals since 
walls there are thick compared with the inverse of the Fermi 
wavelength $k_{F}^{-1}$.
However, the 
most significant feature of a mesoscopic system is the effect of 
the quantum coherence among electrons, which affects substantially the 
low energy transport properties in disordered systems\cite{Bergmann84}.
Interesting point in such weakly localized case is that  
even a small perturbation can 
result in a measurable change in the resistivity of the 
entire sample by disturbing the coherence\cite{Lee85}.
Thus it is natural to expect that  in mesoscopic magnets
the rearrangement of the magnetization 
affects the quantum transport strongly.
In fact it has been predicted that 
in a disordered wire of metallic ferromagnet a domain wall causes a 
dephasing among electrons and thus decreases the quantum correction 
to the resistivity, in contrast to the contribution to 
the classical resistivity\cite{TF97}.

In this paper we will study theoretically   
the transport properties of a non-magnetic conduction layer
sandwiched between two ferromagnetic layers as shown in 
Fig. \ref{FIGmultilayers}, where the $z$-axis is chosen 
perpendicular to the layer.
Both the magnetization and the current are assumed to lie in the 
plane in $x$-direction. (Even if the magnetization is perpendicular to 
the current, the following result is not changed.) 
The calculation is based on the linear response theory.
We assume that the metallic layer is disordered and the resistivity is 
dominated by the normal impurities, thus treating the effect of magnetic 
layers perturbatively.
We assume $d \gg \ell \gg k_{F}^{-1}$, where $d$ is the thickness of 
the conduction layer and $\ell$ is the elastic mean free path.
(We neglect the spin dependence of $k_{F}$.)
The conduction electron feels a magnetic field 
at the interface with ferromagnetic layers.
Within the classical argument of resistivity 
this magnetic field does not affect the in-plane resistivity 
in the case of an ideally flat interface we consider.
It turns out, however, that it affects the quantum correction to 
the resistivity if the two spin channels are mixed by the 
spin flip scattering. 
As source of spin flip scattering we include the spin-orbit (SO) 
interaction, which is known to affect 
the quantum correction  at low temperature in, for 
example, Cu film\cite{Kobayashi80,Maekawa81}. 
The case of isotropic SO interaction is considered.
We consider a thin layer (typically $d\lesssim 2-300\AA$) and thus 
neglect the effect of the orbital motion due to the internal field.

The effect of the magnetic layers is represented by the interaction
\begin{equation}
H_{\rm int}=-\mu_{B} \int d^3 x h(z) c^\dagger \sigma_{z} c,
\label{Hint}
\end{equation}
where $h(z)$ is the magnetic field supposed to simulate the local 
field at the interface with the ferromagnetic layers 
and $\mu_{B}$ is the Bohr magneton.
The quantization ($z$-) axis of electron spin is chosen along the direction 
of the magnetic field (i.e., spatial $x$-axis).
The conductivity is evaluated 
from the current-current correlation function, and the interaction 
eq. (\ref{Hint}) is treated perturbatively to the second order. 
In the classical transport theory the contribution is a made up of a 
self-energy (SE) and a vertex correction (VC) type processes, but 
these two processes cancel each other in the case of flat interface 
because of the symmetry.
This interaction, however, has a finite effect on the quantum 
correction to  
the conductivity, since it modifies
the coherence of the electron wave function.
The effect on the quantum correction would be discussed most 
conveniently in terms of the Cooperon (particle-particle ladder
), which represents the enhancement of the backscattering amplitude 
due to the coherence\cite{Bergmann84}.
The conductivity correction is expressed in terms of the correction 
to the Cooperon, $\delta\Gamma$, diagrammatically as in Fig. 
\ref{FIGnoSO} (a) (see also eq. (\ref{sigmaQ})).\cite{delgammapert}
In calculation of the quantum correction we neglect the quantity of 
$o(k_{F}\ell)^{-1}$.

First we consider the case without the
SO interaction. 
In this case only $\delta\Gamma$ with $\sigma'=\sigma$ 
(Fig. \ref{FIGnoSO}(a)) contributes. This contribution 
is made up of two processes of the 
SE and VC (Fig. \ref{FIGnoSO}(b))
(Note that $h(z)$ is static).
The bare Cooperon here (denoted by shaded line), 
with the momenta of the two incoming electrons $k$ 
and $-k+q$ ($|k|\simeq k_{F}$)) behaves 
at $q\lesssim \ell^{-1}$  as $\Gamma_{0}(q)=(Dq^{2}\tau+\kappa)^{-1}$, 
where 
$D\equiv \hbar^{2}k_{F}^{2}\tau/3m^{2}$ is the diffusion constant, 
$\tau$ being 
the elastic lifetime ($\ell= \hbar k_{F}\tau/m$), 
$\kappa\equiv \tau/\tau_{\varphi} \ll1$, 
$\tau_{\varphi}$ being the inelastic lifetime, and
$N(0)$ is the density of states at the Fermi energy.
We consider below the most important contribution, which 
comes from the region $|q|, |Q|\lesssim \ell^{-1}$
($Q$ being the momentum transfer 
due to the interaction (\ref{Hint}), which has only $z$-component).
It is easy to see then that the two processes cancels with each other.
In fact the summation over the electron momentum $\mbox{\boldmath{$k$}}'$ 
and $\mbox{\boldmath{$k$}}''$ 
in the SE and VC-type gives rise to $(I_{qQ})^2$ and $|I_{qQ}|^2$, 
respectively, where
$I_{qQ}\equiv \sum_{\mbox{\boldmath{$k$}}}G^{+}_{\mbox{\boldmath{$k$}}}
G^{+}_{\mbox{\boldmath{$k$}}+Q}G^{-}_{-\mbox{\boldmath{$k$}}+\mbox{\boldmath{$q$}}}$.
Here $G_{\mbox{\boldmath{$k$}}}^{\pm}$ is the electron Green function; 
$G_{\mbox{\boldmath{$k$}}}^{(\pm)}\equiv 
1/(\pm i\hbar/2\tau-\epsilon_{\mbox{\boldmath{$k$}}})$ 
($\epsilon_{\mbox{\boldmath{$k$}}}\equiv 
\hbar^{2}\mbox{\boldmath{$k$}}^{2}/2m-\epsilon_{F}$).
The sign $\pm$ corresponds to the sign of Matsubara frequency, 
and the difference of SE and VC contribution is due to the difference 
of this sign.
Since $I_{qQ} \simeq -2\pi iN(0)\tau^{2}/\hbar^2$ is pure imaginary, the 
contribution from 
the two processes cancels each other.
Hence the conductivity is not 
affected by the magnetic layers in the absence of the SO interaction.

Now we include the SO interaction.
The spin conserving process considered in Fig. \ref{FIGnoSO}(b) 
vanishes due to the same reason as before.
In contrast the correction to a Cooperon with a spin flip
($\sigma'=-\sigma$) ($\Gamma_{+-}$ in Fig. \ref{FIGqc}(a))  
has a finite effect\cite{gamma0}. 
This correction ($\equiv 
\delta\Gamma_{+-}$) is shown in Fig. \ref{FIGqc}(b).
Other processes with two or less number of Cooperons gives smaller 
contributions for small $\kappa$.
Here the bare Cooperon $\Gamma_{+-}$ 
 is proportional to the strength of the SO interaction, and 
is calculated as\cite{Maekawa81}
\begin{equation}
	\Gamma_{\rm +-}(q)=-\frac{2\alpha} 
	{(Dq^{2}\tau+\kappa)(Dq^{2}\tau+\kappa+4\alpha)},
\end{equation}
where $\alpha \equiv \tau/3\tau_{\rm so}\ll1$, $\tau_{\rm so}$ 
being the inelastic lifetime due to SO interaction.
The other Cooperon in Fig. \ref{FIGqc}(a), $\Gamma_0'$, 
is a $\Gamma_0$ modified by the SO interaction;
\begin{equation}
	\Gamma_{\rm 0}'(q)=\frac{1} {(Dq^{2}\tau+\kappa)} 
	\frac{Dq^{2}\tau+\kappa+2\alpha}{(Dq^{2}\tau+\kappa+4\alpha)}.
\end{equation}
In processes in Fig. \ref{FIGqc}(b) 
the cancellation between SE and VC does not occur, because of the 
different signs arising from 
the two interaction vertices ($\propto \sigma_{z}$), and they give 
the equal contribution.
In fact the sum of the first and second processes 
 ($\equiv \delta \Gamma_{\rm i,ii}$)
is calculated as (factor of 2 is from the complex conjugate process)
\begin{eqnarray}
	\delta\Gamma_{\rm i,ii}(q) & = & -2\sum_{Q} \left(\frac{\hbar}{2\pi 
	N(0)\tau}\right)^{3}(\Gamma_{+-}(q))^{2}\Gamma_{+-}(q+Q) 
	|h(Q)|^{2} \left[ (I_{qQ})^{2}- |I_{qQ}|^{2} \right]
	\nonumber  \\
	 & = & 4\sum_{Q} \frac{\tau}{2\pi 
	N(0)\hbar}(\Gamma_{+-}(q))^{2}\Gamma_{+-}(q+Q) 
	|h(Q)|^{2},
	\label{delGam1}
\end{eqnarray}
where $h(Q)$ is the Fourier transform of $h(z)$, 
and $[2\pi N(0)\tau/\hbar]^{-1}$ stands for the strength of the 
impurity scattering.
In this way we obtain the correction to $\Gamma_{+-}$  as
\begin{equation}
		\delta\Gamma_{+-}(q) = 4\sum_{Q} \frac{\tau}{2\pi 
	N(0)\hbar}|h(Q)|^{2} C(\mbox{\boldmath{$q$}},Q),\label{delGam}
\end{equation}
where
\begin{equation}
	C(\mbox{\boldmath{$q$}},Q)\equiv 
	(\Gamma_{+-}(q))^{2}\Gamma_{+-}(q+Q) +\Gamma_{0}'(q) 
	\left\{ 
 \Gamma_{0}'(q) \Gamma_{+-}(q+Q)-2\Gamma_{+-}(q)\Gamma_{0}'(q+Q)
 \right\} .	
\end{equation}
Each term here corresponds to the diagram i-ii), iv) and iii),v), 
respectively.
By use of this expression
the quantum correction to the conductivity induced by the magnetic 
layers is obtained as (see Fig. \ref{FIGnoSO}(a)) (including a factor 
of 2 coming from spin)
\begin{eqnarray}
        \delta\sigma& =& \frac{2\hbar}{2\pi}
        \left(\frac{e\hbar\mu_{B}}{m}\right)^{2}
        \frac{1}{V}
        \sum_{\mbox{\boldmath{$k$}}}\sum_{q\lesssim\ell^{-1}} 
         k_{x}(-k+q)_{x} G^{+}_{\mbox{\boldmath{$k$}}}G^{-}_{\mbox{\boldmath{$k$}}}
	G^{+}_{-\mbox{\boldmath{$k$}}+\mbox{\boldmath{$q$}}}
	G^{-}_{-\mbox{\boldmath{$k$}}+\mbox{\boldmath{$q$}}} 
	\delta\Gamma_{+-}(q)     \nonumber\\
	&\simeq&    \frac{8e^{2}}{3\pi\hbar} 
	\left(\frac{\mu_{B}\tau \ell}{\hbar}\right)^{2}
       \sum_{Q\parallel z, Q\lesssim\ell^{-1}} |h(Q)|^{2} 
       \frac{1}{V}\sum_{q\lesssim\ell^{-1}} C(\mbox{\boldmath{$q$}},Q).
	\label{sigmaQ}  
\end{eqnarray}

The effect of the magnetic layers becomes most significant when the 
electron coherence is kept throughout the layer thickness, i.e., 
$d \ll \ell_{\varphi}$ ($\ell_{\varphi}\equiv\ell/\sqrt{3\kappa}$ 
being the inelastic mean free path). 
In this case the system behaves as in two-dimensions from the point of 
view of the coherence, namely, only the 
$Q=0$ and $q_z=0$ component become important and thus $h(Q)$ 
becomes essentially a uniform magnetic field.
In this case eq. (\ref{sigmaQ}) reduces to
\begin{eqnarray}
\delta\sigma &=&-\frac{16e^{2}}{3\pi\hbar} 
\left(\frac{\mu_{B}h(Q=0)\tau}{\hbar}\right)^{2} \ell^2 \alpha 
\frac{1}{V} \sum_{q_x,q_y} 
 \frac{1}{(Dq^2 \tau+\kappa)^2 (Dq^2 \tau+\kappa+4\alpha)^2 }
\nonumber\\
&=& 
-\frac{4e^{2}}{3\pi^{2}\hbar} 
\left(\frac{\mu_{B}h(0)\tau}{\hbar}\right)^{2} 
\frac{\alpha F(\kappa,\alpha)}{d\kappa^3},
\label{delsig1}
\end{eqnarray}
where
\begin{equation}
	F(\kappa,\alpha)\equiv \frac{3}{16} \frac{\kappa^3}{\alpha^2} 
	\left[\frac{1}{\kappa}+\frac{1}{\kappa+4\alpha} 
	-\frac{1}{2\alpha}\ln\left(1+\frac{4\alpha}{\kappa}\right) \right].
\end{equation}
In the case of weak SO interaction ($\alpha\ll\kappa$), 
$F(\kappa,\alpha)=1+O(\alpha/\kappa)$.

To proceed further we need the explicit profile of $h(Q)$. 
We consider two cases where the magnetization of the two ferromagnetic 
layers are parallel (P) or 
anti-parallel (AP) to each other (Fig. \ref{FIGmultilayers}).
Choosing $z=0$ as the center of the conduction layer, we assume the effective 
magnetic field at the interface is written as
 \begin{equation}
 	h^{\pm}(z)=h_{0}a(\delta(z-d/2)\pm\delta(z+d/2)),
 	\label{hdef}
 \end{equation}
for P and AP cases (denoted by $h^{+}$ and $h^{-}$, 
respectively) ($h_{0}$ is a constant which represents the strength of the 
field and $a$ is the scale of the penetration of the effective 
field, which is of order of a lattice constant).
The Fourier transform of $h^{\pm}(z)$ are obtained as 
$h^{+}(Q)\equiv(1/d)\int_{-d/2}^{d/2} h^+(z)e^{-iQz}dz
=(h_{0}a/d)\cos({Qd/2})$ and 
$h^{-}(Q)=-i(h_{0}a/d)\sin(Qd/2)$, 
where $Q$ takes values of $Q=\pi n/d$, $n$ 
being an integer from $-N/2$ to $N/2$ ($N\equiv d/a\gg1$ is the number of 
atomic layers). 
The magnitude of the MR due to the flip of the 
magnetization is then written as 
$\Delta\rho/\rho_{0}=-(\delta\sigma^{+}-\delta\sigma^{-})/\sigma_{0}$, 
where $\delta\sigma^{\pm}$ denotes the quantum correction for the 
configuration $h^{\pm}$ and 
$\rho_{0}=\sigma_{0}^{-1}\equiv 6\pi^{2}(\hbar/e^{2}k_{F}^{2}l)$ is 
the resistivity due to impurities.

In the case of $d \ll \ell_{\varphi}$ considered in eq.(\ref{delsig1}), the 
MR close to the switching field is obtained as 
\begin{equation}
\frac{\Delta\rho}{\rho_{0}}
	= 2
	\left(\frac{\Delta_{0}}{\epsilon_{F}}\right)^{2}
	\left(\frac{a}{d}\right)^{2} \frac{\ell}{d} \frac{\alpha}{\kappa^3} 
	F(\kappa,\alpha) 
,	\label{Deltarhoratio}
\end{equation}
where $\Delta_{0}\equiv \mu_{B}h_{0}$ is the Zeeman 
splitting at the interface, and $\Delta_{0}(a/d)$ is a measure of the 
averaged splitting.

If the magnetization of the two 
ferromagnetic layers are AP to each other in the absence 
of an external magnetic field (as is realized by controlling 
$d$\cite{Bruno91}), positive $\Delta\rho$ obtained here 
contributes to a positive MR close to a switching field, 
where the magnetization flips.
The MR in real experiments is affected also by the interface 
roughness and spin-dependent scattering there.
The effect of the dephasing considered here
will be separated from such effects 
by looking into the temperature dependence of $\Delta\rho$.
In fact as the temperature is lowered $\kappa$ decreases since 
dephasing due to phonons and the electron-electron 
interaction are suppressed (by power law $\kappa \propto T^{p}, 
p\sim O(1)$\cite{Bergmann84}). 
Then $\Delta\rho$ due to the dephasing will become large
according to eq. (\ref{Deltarhoratio}), 
while the resistivity change due to other classical origins would not change 
so much at very low temperature.

Consider a layer of $d= 100\AA$ and $\ell=30\AA$, and take the
effective exchange splitting of the $s$ electron
at the interface as $\Delta_{0}/\epsilon_{F}\sim 3\times 10^{-2}$. 
Then if $\kappa=10^{-2}$, which means that the inelastic 
diffusion length is 
$\ell_{\varphi}\sim 5.7\ell$, the system behaves as in two-dimensions. 
If $\alpha/\kappa\sim 0.5$ then we obtain $\Delta\rho/\rho_{0}=0.5\times 
10^{-3}$.
For a material with a larger induced Zeeman splitting, $\Delta_{0}$, we 
expect a bigger effects.

Let us discuss the case where the conduction layer is a soft 
ferromagnet which is in contact with a hard ferromagnet as realized 
by use of NiFe/CoSm\cite{Mibu98}.
In such system the magnetization at the interface is fixed by a hard 
magnet at small field. 
Thus an artificial twisted structure of the magnetization similar to a 
domain wall can be formed inside the soft magnetic layer by applying 
a magnetic field.
Although the configuration of magnetization differs from the case of 
non-magnetic layer in contact with ferromagnets considered above, 
the argument goes parallel.
In fact by use of a local gauge transformation\cite{TF97}, the 
correction to the in-plane conductivity turns out to be obtained by 
eq. (\ref{sigmaQ}) with the replacement of 
 $|\mu_{B}h(Q)|^{2} \rightarrow 
(1/12)(\hbar^{2}k_{F}/m)^{2}|a(Q)|^{2}$, where $a(Q)\equiv 
(1/d)\int dz e^{iQz}\nabla_{z}\theta(z)$, $\theta(z)$ being the angle 
which describes the direction of the magnetization.
(Here we assumed $\Delta\tau/\hbar \ll 1$, $\Delta$ being the Zeeman 
splitting in the soft magnetic layer).
If $\theta$ changes uniformly from $0$ to $\theta_{0}$ inside a conduction 
layer, i.e., $\nabla_{z}\theta=\theta_{0}/d$, then the quantum 
correction in the case of $d \ll \ell_{\varphi}$ is obtained 
(for $\alpha \ll\kappa$) as 
\begin{equation}
	\frac{\Delta\rho}{\rho_{0}}
	  =\frac{2\theta_{0}^{2}}{3}
	\frac{\ell}{k_{F}^{2} d^{3}} \frac{\alpha}{\kappa^{3}}.
	\label{delrhoF}
\end{equation}
This is positive, and increases for larger twist angle, $\theta_{0}$.
A measurement on NiFe(300\AA)/CoSm\cite{Mibu98} indicated 
an increase of the resistivity after subtracting the anisotropic 
MR as $\theta_{0}$ increases.
The change was about 0.08\%  of the total resistivity 
($\sim9.6\mu\Omega$cm) at 5K. 
In Ref. \cite{Mibu98}, the effect of a strongly 
spin-dependent lifetime in ferromagnets is suggested as a 
possible origin of the observed $\Delta\rho$.
However, as eq. (\ref{delrhoF}) indicates, the dephasing due to 
twisted magnetization may contribute to positive $\Delta\rho$ at low 
temperature.
In fact if the inelastic mean free path is not very long (e.g., 
$\kappa\sim0.1$) (the mean free path is estimated as $\ell\sim30\AA$)
the expected effect due to dephasing is 
$\Delta\rho/\rho_{0}=0.08$\% for $\alpha/\kappa\sim0.5$.
In Ref. \cite{Mibu98} increase of $\Delta\rho$ has been observed as 
the temperature is lowered to 2K, although $\Delta\rho$ itself still exists 
at high temperature of 100K. 
This enhancement at low temperature might be due to the dephasing.

In conclusion we have studied based on the linear response 
theory the effects of the magnetic layers on the 
in-plane resistivity of a disordered conduction layer 
sandwiched between two ferromagnetic layers with ideally flat interface.
It has been shown that while the magnetization at the interface with 
the magnetic layers does not affect 
the classical resistivity, it affects the quantum correction to 
the resistivity at low temperature in the presence of the spin-orbit 
interaction, resulting in a larger resistivity for the 
parallel configuration of the 
magnetization of the two ferromagnetic layers.

The authors thank K. Mibu for valuable discussions. 
This work was partially supported by a Grand-in-Aid for Scientific 
Research on the Priority Area ``Nanoscale Magnetism and Transport''  
(No.10130219)
and ``Spin Controlled Semiconductor 
Nanostructures'' (No.10138211) from the Ministry of Education, Science, 
Sports and Culture.
%



%
\begin{figure}[htb]
\caption{A layer of a non-magnetic metal with ferromagnetic boundaries with 
two configurations of the magnetization.
\label{FIGmultilayers}}
\caption{(a): The correction to conductivity expressed by the Cooperon, 
$\delta\Gamma$. The electron spin is 
denoted by $\sigma$ and $\sigma'$.
(b): Corrections to the Cooperons by the ferromagnetic boundaries but
in the absence of 
spin-orbit interaction (i.e., $\sigma=\sigma'$).
Two electron lines carry Matsubara frequencies of different sign.
Cooperons in the absence of ferromagnetic boundaries are denoted by 
shaded lines and wavy lines represent the 
interaction with the boundary ferromagnets, eq. (\protect\ref{Hint}).
Two of the bare Cooperons carry a momentum of $q$ and one carries $q+Q$.
\label{FIGnoSO}}
\caption{(a): Cooperons with spin-flip, 
$\Gamma_{+-}$, and $\Gamma_{0}'$ in the presence of spin-orbit 
scattering. 
(b):Dominant corrections to the Cooperons with a spin flip, 
$\delta\Gamma_{+-}$. Diagrams iii)-v) differs in the way of sequence of 
$\Gamma_{+-}$ and $\Gamma_0'$s, and thus the spin indices of the 
internal electron lines differ.
\label{FIGqc}}
\end{figure}
%

\begin{thebibliography}{99}
\bibitem{delgammapert}
If the spatial dependence of $h(z)$ is neglected, the calculation 
of $\delta\Gamma$ here corresponds to a 
perturbative expansion of the full Cooperon calculated in 
Ref. \protect\cite{Maekawa81} with respect to the Zeeman 
splitting.
\bibitem{gamma0}
The correction to $\Gamma_{0}'$ does not contribute to the 
conductivity because of the spin conservation at the current vertex.
\end{thebibliography}
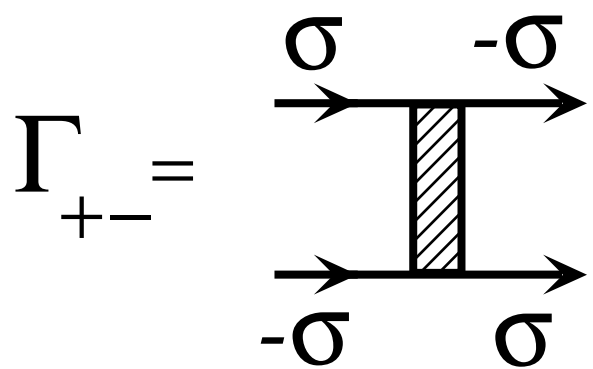
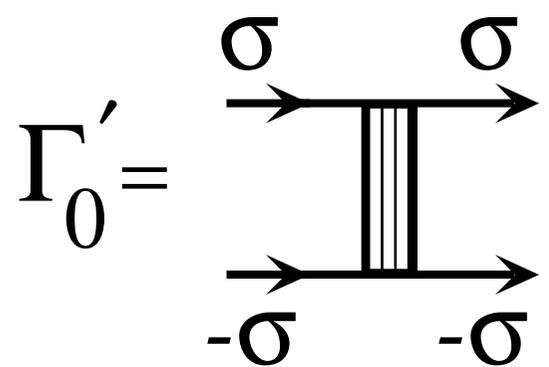
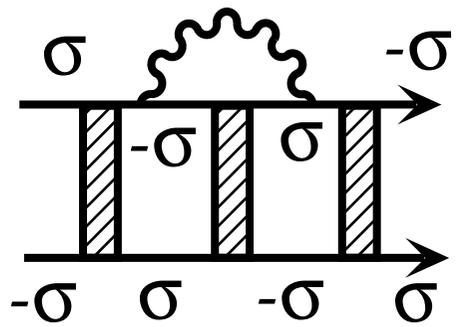
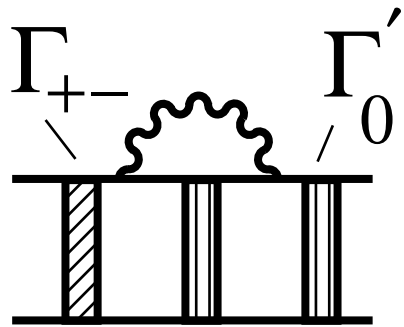
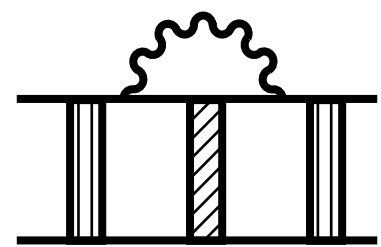
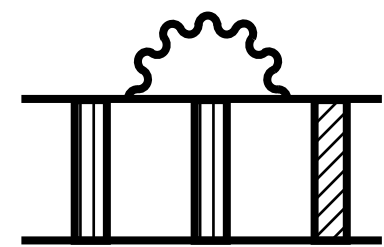
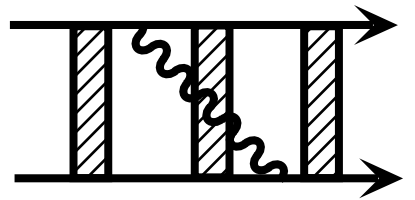
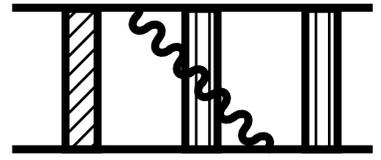
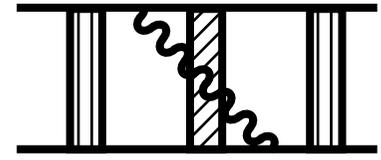
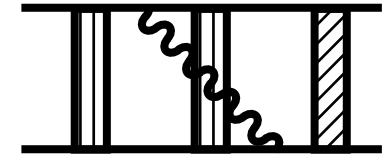
\end{document}